\documentclass[amsmath,amssymb,preprint]{revtex4-1}
\usepackage{graphicx}
\usepackage{epsfig}
\usepackage{dcolumn}
\usepackage{color}
\usepackage{ulem}
\usepackage{subcaption}
\usepackage{amsmath}
\usepackage{tabularx}
\usepackage{array}

\begin{document}

\title{Properties of a static dipolar impurity in a 2D dipolar BEC}
\author{Neelam Shukla} 
\author{Jeremy R. Armstrong}

\affiliation{Department of Physics \& Astronomy, University of Nebraska at Kearney, NE-68849 USA}

\begin{abstract}
  We study a system of ultra cold dipolar Bose gas atoms confined in a two-dimensional (2D) harmonic trap with a dipolar impurity implanted at the center of the trap.  Due to recent experimental progress in dipolar condensates, we focused on calculating properties of dipolar impurity systems that might guide experimentalists if they choose to study impurities in dipolar gases.  We used the Gross-Pitaevskii formalism solved numerically via the split-step Crank-Nicolson method. We chose parameters of the background gas to be consistent with dysprosium (Dy), one of the strongest magnetic dipoles and of current experimental interest, and used chromium (Cr), erbium (Er), terbium (Tb), and Dy for the impurity.  The dipole moments were aligned by an external field along what was chosen to be the $z$-axis, and studied 2D confinements that were perpendicular or parallel to the external field.  We show density contour plots for the two confinements, 1D cross sections of the densities, calculated self-energies of the impurities while varying both number of atoms in the condensate and the symmetry of the trap. We also calculated the time evolution of the density of an initially pure system where an impurity is introduced.  Our results found that while the self-energy increases in magnitude with increasing number of particles, it is reduced when the trap anisotropy follows the natural anisotropy of the gas, i.e., elongated along the $z$-axis in the case of parallel confinement.  This work builds upon work done in Bose gases with zero-range interactions and demonstrates some of the features that could be found when exploring dipolar impurities in 2D Bose gases.    
\end{abstract}

\maketitle

\section{Introduction}
Dipolar BECs have long been of interest in cold atom physics due to the large variety of physics they display~\cite{lahaye, chomaz}.  The first such system was a condensate of $^{52}$Cr atoms discovered in 2004~\cite{griesmaier} utilizing its magnetic dipole moment, but BECs using electric dipole moments have recently been realized~\cite{bigagli}.  While the condensates themselves are of interest, experiments with impurity species implanted in them may become of great interest in the coming years.  Boson impurities in non-dipolar condensates have been well-studied both experimentally~\cite{hu2016, jorgensen} and theoretically~\cite{scazza, mistakidis,grusdt2024}.

Given the high degree of experimental control and precision, cold atomic systems are a great sandbox for impurity physics.  When an impurity is introduced into a cold atomic system, it becomes \textit{dressed} by the medium particles.  These interactions alter the properties of the impurity from its bare form and also allow the impurity to act as a local probe of its environment ~\cite{volosniev2019solvable, mehboudi,bouton2020}.  The impurity may also alter properties of the bath local to the impurity, as it will change the bath density around it.  The extent of its effects depend on both the strength and range of the impurity's interactions and the properties of the medium.  These effects have been studied experimentally and have also been studied in the case of two-dimensional (2D) systems~\cite{zhang2012,ong2015spin,koschorreck}.

There has been other work with impurities in dipolar media \cite{Kain2014,Ardila2018,Volosniev2023,shukla}.  Most of these works \cite{Kain2014,Ardila2018,Volosniev2023} have been in momentum space, while our work is in coordinate space.  This gives us access to important quantities such as the density.  Our aim is to further the exploration of impurities in cold dipolar gases with a study in which the impurity is also a dipole.  Furthermore, the gas is confined to two spatial dimensions, which has been less explored than 3D.  This is in contrast to our previous work, \cite{shukla}, where we worked in 3D and with a Gaussian impurity which was strongly interacting and had other tuneable properties.  In our system, the dipolar BEC is confined in a 2D harmonic trap with a dipolar impurity implanted inside the trap.  The dipoles are polarized in the $z$-direction, and we have studied both of the extreme orientations, where the trap contains the dipoles in the plane perpendicular to the polarization direction or parallel to it. 

This paper is structured as follows: after the introduction, we will discuss the methods used in our calculations, we will then report our results for the two geometries, discuss them, and then conclude.

\section{Methods}

Our dipolar BEC consists of $N$ dipoles in a 2D harmonic trap with a single impurity dipole implanted inside.  The Hamiltonian of our system is:
\begin{eqnarray}
  H&=&\sum_{k=1}^N\left(\frac{p_k^2}{2m}+V_{trap}(\mathbf{r}_k)+\beta V_{dip}(\mathbf{r_k}-\mathbf{r})\right)+\frac{p^2}{2m_{imp}}\nonumber\\&&+\sum_{i<j}^NV_c(\mathbf{r_i}-\mathbf{r_j})
  +\sum_{i<j}^NV_{dip}(\mathbf{r_i}-\mathbf{r_j}),
 \label{hamil} \end{eqnarray}
where $m$ is the mass of a dipolar boson, $m_{imp}$ and $\mathbf{r}$ are the mass and position of the impurity dipole, respectively, $V_c$ is the contact interaction, $V_{dip}$ is the dipole-dipole interaction, which will be detailed subsequently, and $\beta$ is a scaling factor for the strength of the dipole-impurity interaction.  

The contact interaction is $V_c(\mathbf{r})=4\pi\hbar^2a\delta(\mathbf{r})/m$, where $a$ is the scattering length.  The dipole-dipole interaction is long-range and anisotropic.  In general form, it reads for magnetic dipoles
\begin{equation}
V_{dip}(\mathbf{r})=\frac{\mu_0}{4\pi}\frac{d^2-3(\mathbf{d} \cdot \mathbf{\hat{r}})^2}{|\mathbf{r}|^3},\label{OGdipole}
\end{equation}
where, $\mathbf{d}$ is the dipole moment, $\mu_0$ is the permeability of free space, and $\mathbf{\hat{r}}$ is the unit vector along $\mathbf{r}$.  
In our system, the dipoles are polarized in the $z$-direction by an external field.  In this case, the dipole-dipole interaction can be written as
\begin{equation}
V_{dip}(\mathbf{r})=\frac{C_{dd}}{4\pi}\frac{1-3\cos^2(\theta_d)}{r^3}, \label{dipole}
\end{equation}
where, $\theta_d$ is the angle between $\mathbf{r}$ and $\mathbf{d}$.  We have used the standard notation $C_{dd}=\mu_0d^2$, allowing us to introduce a useful length scale $a_{dd}=C_{dd}m/(12\pi\hbar^2)$, known as the 'dipolar length'\cite{lahaye}.  The dipolar length allows us to easily compare the relative strengths of interactions in the problem, by comparing it with the scattering length of the contact interaction.  The scaling factor that appears in Eq. (\ref{hamil}) allows us to treat an impurity with a different dipole moment than the background dipolar gas.  In such a situation, Eqs. (\ref{OGdipole}) and (\ref{dipole}) would need two different dipole moments, $\mathbf{d_{g}}$ and $\mathbf{d_{i}}$, instead of the $d^2$ terms.  We have chosen to handle this by using a scaling factor $\beta=|\mathbf{d_i}|/|\mathbf{d_g}|$, where $\mathbf{d_g}$ is the dipole moment of a bulk gas atom and $\mathbf{d_i}$ is the dipole moment of the impurity.  In terms of the dipolar lengths $a_{dd,g}$ and $a_{dd,i}$ (the dipolar lengths of the gas and impurity, respectively), $\beta=\sqrt{\frac{a_{dd,i}m_g}{a_{dd,g}m_i}}$, where $m_g$ and $m_i$ are the masses of a dipolar gas atom and impurity atom, respectively.  Note that the dipole-dipole interaction is singular at the origin.  In order to treat this difficulty, at the origin we have taken the potential to be a constant value equal to the average of the potential value at the four nearest grid points to the origin in our numerical solution.  The short-range behavior of the dipole-dipole interaction is beyond the scope of this paper and not vital to our findings.

In order to solve Eq. (\ref{hamil}), we assume that our wave function $\psi$, can be written as a product of one-particle wave functions, i.e., $\psi(\mathbf{r_1},\mathbf{r_2},\dots,\mathbf{r_N})=\psi(\mathbf{r_1})\psi(\mathbf{r_2})\dots\psi(\mathbf{r_n})$.  If there was no trap, we can transform the resulting Schr\"odinger equation ($H\psi=E\psi $) into the frame of the impurity and obtain~\cite{gross1962,Volosniev2017,Hryhorchak_2020,Jager2020,Enss2020,Guenther2021}:
\begin{eqnarray}
  -\frac{1}{2}\left(\frac{1}{m}+\frac{1}{{m_{imp}}}\right)\nabla^2\psi(\mathbf{r})+\beta V_{dip}(\mathbf{r})\psi(\mathbf{r})
  +\psi(\mathbf{r})\int\left[V_c(\mathbf{r'}-\mathbf{r})+V_{dip}(\mathbf{r'}-\mathbf{r})\right]|\psi(\mathbf{r'})|^2d\mathbf{r'}\nonumber\\
  =E\psi(\mathbf{r}).\label{GPE}
\end{eqnarray}
This equation is a Gross-Pitaevskii equation (GPE).  For numerical convenience as well as a recognition of experimental realtities, we will solve the GPE in a trap.  If the trap is large, the properties of the impurity are insensitive to it.  It also makes our results somewhat relevant for mobile impurities.  Alternatively, one can use an external laser to pin the impurity at a certain location, such as the center of the trap \cite{catani}, which results in a similar GPE.  We will solve the GPE numerically by adapting a publicly available code~\cite{kumar2015} to our problem.  The code solves the GPE using the split-step Crank-Nicolson method.  

Our system is in a harmonic trap, which we have initially chosen to be radially symmetric: 
\begin{equation}
  V_{trap}(\mathbf{r})=\frac{1}{2}m\left(\omega^2r^2+\omega^2_{3D}x_3^2\right),
\end{equation}
where $r^2=x^2 + y^2$ or $r^2=x^2 + z^2$, depending on the confinement geometry of the system,  $x_3=z$ or $y$ and $\omega_{3D}$ is the confinement frequency in the third dimension, and we will assume $\omega_{3D} \gg \omega$.  In our calculations, we have chosen the oscillator length, $\ell=\sqrt{\hbar/m\omega}$ to be 1 $\mu$m, which is within the range of typical experimental values.  Our dipolar length is 130$a_0$, which is the value for $^{162}$Dy, which has a very large magnetic moment and is of great current experimental interest~\cite{chomaz}.  We have used $a=150a_0$ for the contact interaction scattering length.  While this is close to the background value for Dy, this is also something that is tunable experimentally via Feschbach-Fano resonances~\cite{chin2010}.  It is also important for the stability of dipolar systems that $a>a_{dd}$~\cite{lahaye}.

The reader will have noticed that the Hamiltonian, Eq. (\ref{hamil}), is a three-dimensional equation, whereas we are interested in studying 2D systems.  As hinted at above, we work in quasi-2D, where one of the confining frequencies in the harmonic trap is large and we assume that the particles are in the ground state in that direction, which is a Gaussian wave function.  That direction is then integrated out and the resulting equation is then in 2D~\cite{kumar2015}.

\section{Results}

In this section we will present our results starting with static properties calculated in the two different geometries mentioned in the introduction: confining the particles to the $xy$-plane (perpendicular to the dipole polarization) and the $xz$-plane (parallel to the dipole polarization).  One of the static results we will show in both subsections is the self-energy of the impurity.  The self-energy is defined as
\begin{equation}
  E_{self}=E[\beta]-E[\beta=0],
  \label{selfE}
\end{equation}
where $E[\beta]$ is the energy of the system with an impurity of strength $\beta$.  This quantity is an important characteristic for an impurity system as it indicates how the system's energy is affected by the introduction of an impurity.  Several different impurities were chosen with different strengths: $^{162}$Dy, $^{52}$Cr, and $^{168}$Er.  While at first glance, it may seem that $^{162}$Dy cannot be an impurity in a gas of $^{162}$Dy atoms, the impurity atom can be kept in a different hyperfine state and thus distinct from the background gas. 

\subsection{$xy$-plane}

\begin{figure}[!h]
\centering
\includegraphics[width=0.48\linewidth]{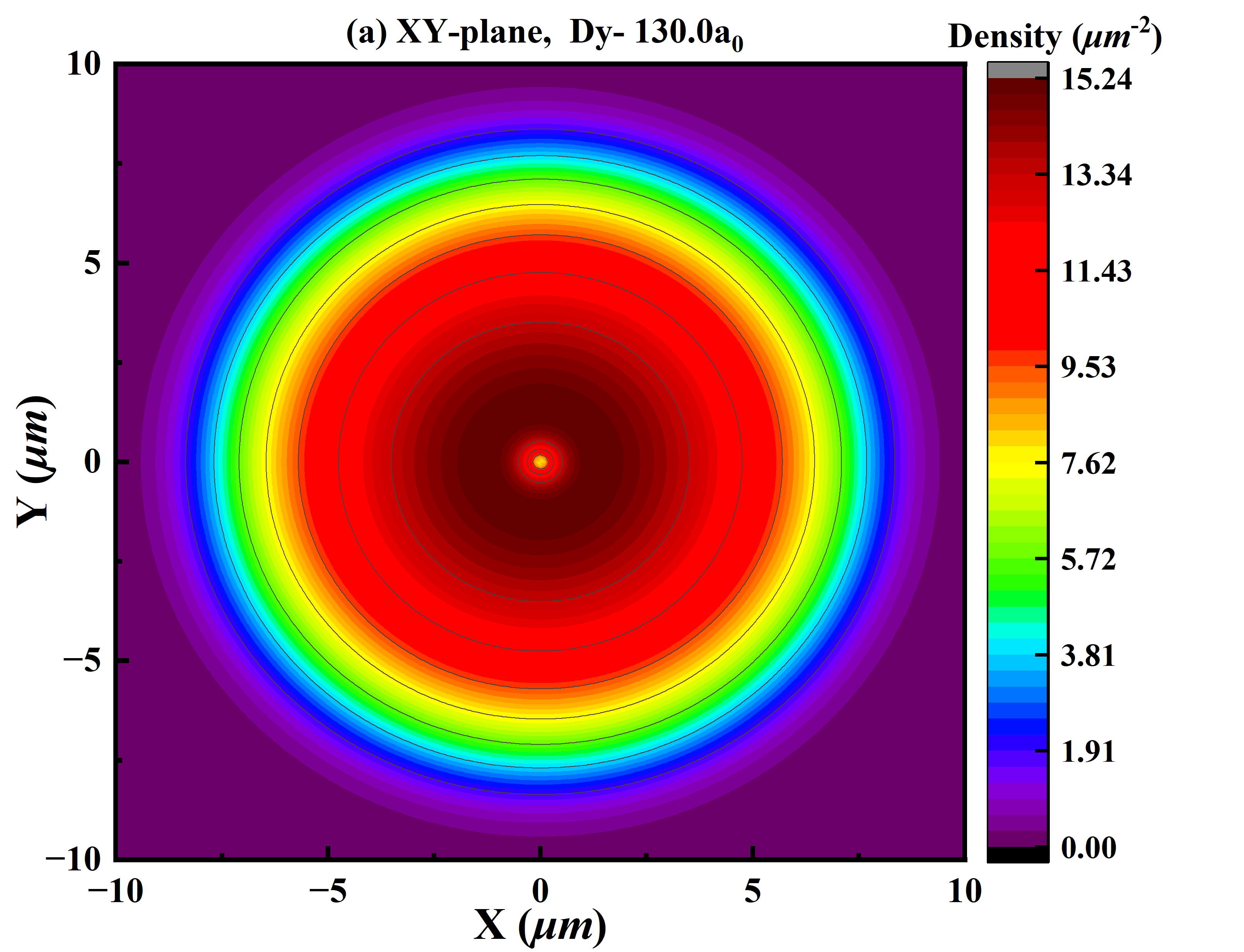}
\includegraphics[width=0.48\linewidth]{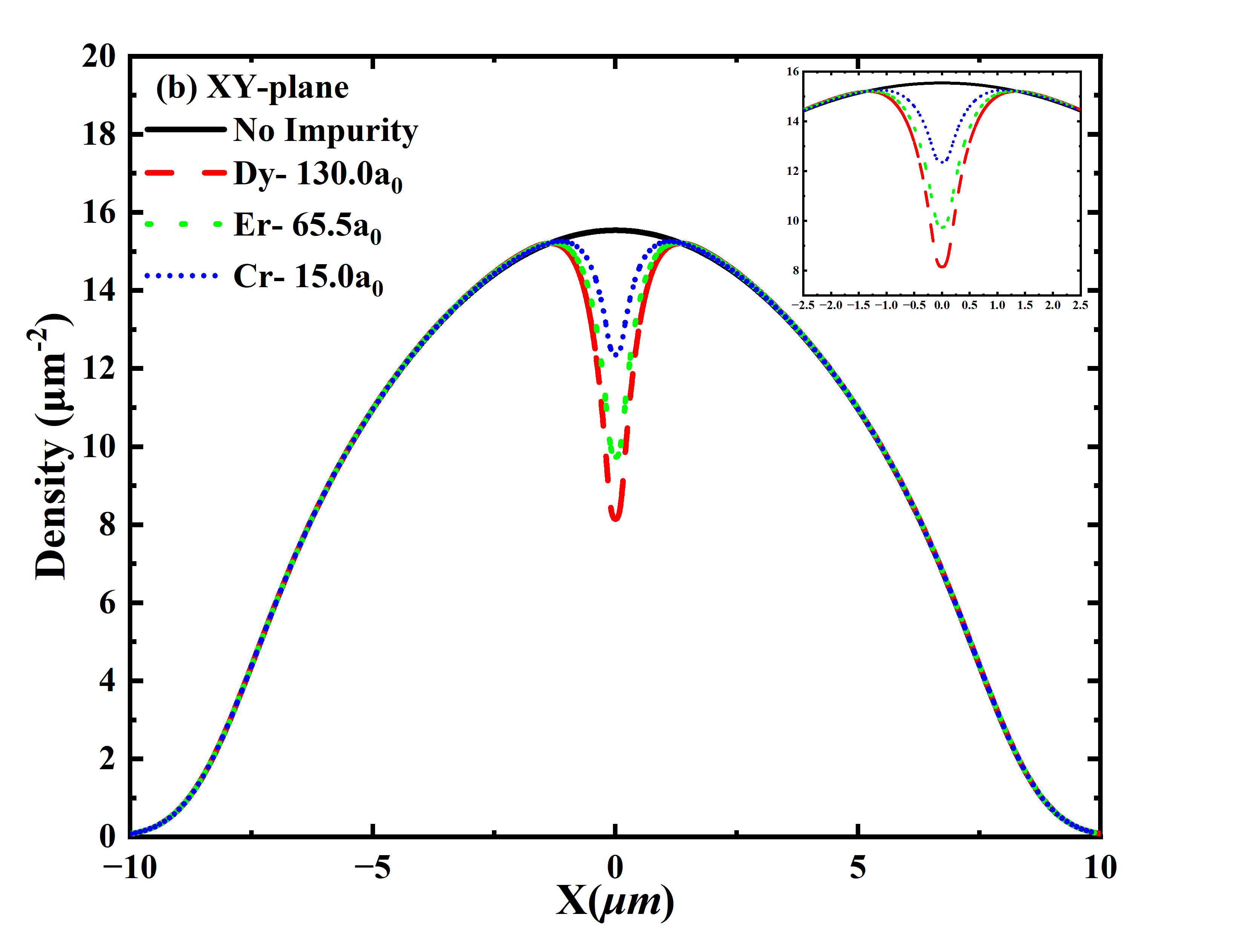}
\caption{\label{fig:XYpannel} Shown in Panel a) is a density contour plot for a gas dipolar gas consisting of 2000 Dy atoms confined in the $xy$-plane with dipole moments polarized perpendicular to the plane with an impurity Dy atom at the origin. Panel b) shows a one dimensional (1D) cross section of the density taken along the $x$-axis for three different impurities as well as the cross section of the system without an impurity. This inset in panel b) shows a close-up of the profile near to the origin. The $y$-axis cross section is not shown as the system is isotropic and thus it is identical to the $x$-axis cross section.}
\end{figure}

\begin{figure}[!h]
\centering
\includegraphics[width=0.48\linewidth]{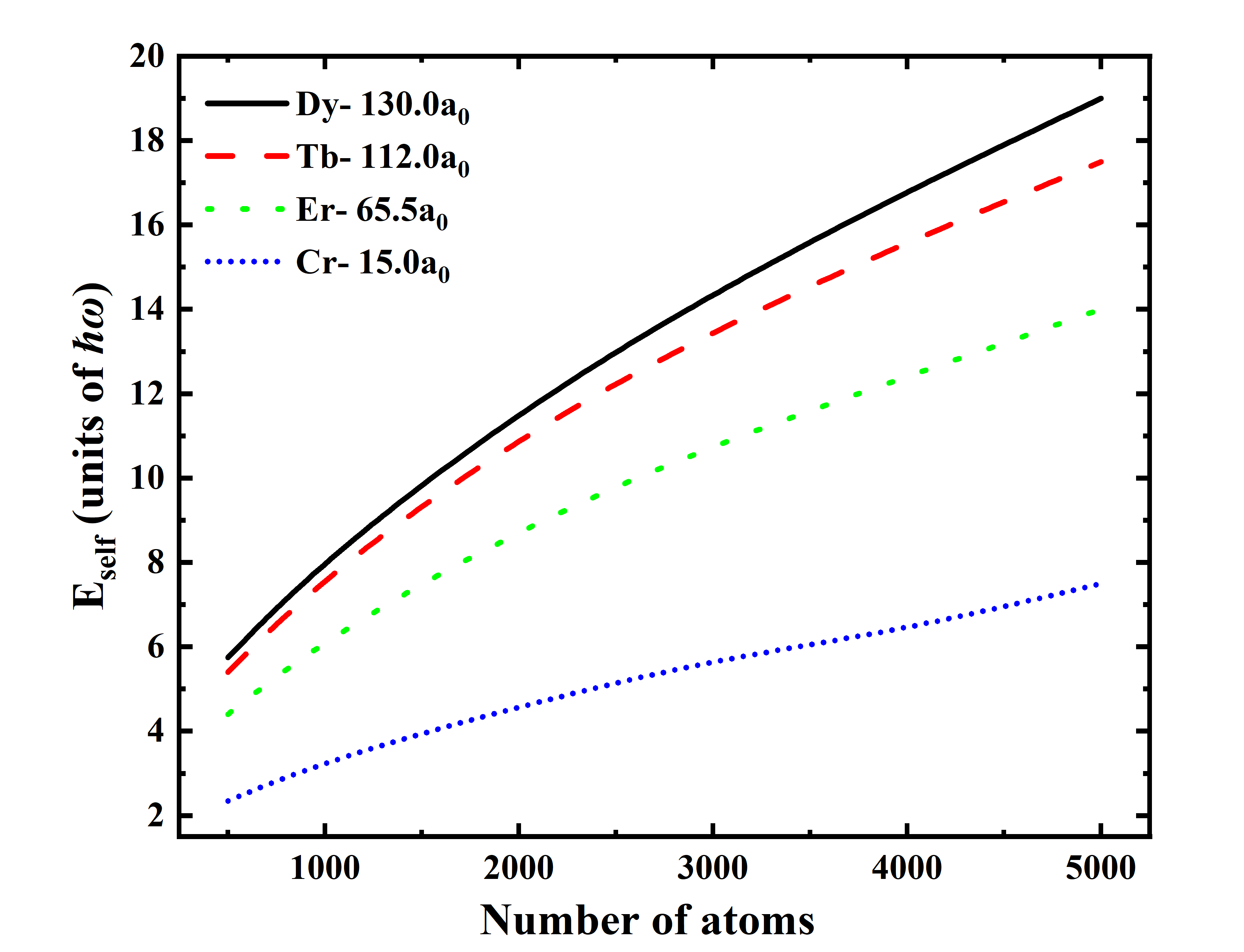}
\caption{\label{fig:XYenergy} Self-energy, Eq. (\ref{selfE}), of an impurity implanted into a dipolar gas confined to the $xy$-plane as a function of the number of particles in the dipolar gas for four different impurities.}
\end{figure}

In the $xy$-plane, the angle $\theta_d$ in Eq. (\ref{dipole}) is $\pi/2$ always, which results in a dipolar interaction that is purely repulsive and isotropic.  Thus, we have a system with only repulsive interactions present that is confined by a harmonic trap.  In panel a) of Figure \ref{fig:XYpannel}, we can see the density contours of this system.  The repulsive impurity at the origin depletes the density there which then concentrates in an annulus around the center.  The contours are circular in shape, fitting with the isotropic nature of the interactions.  In panel b) of Figure \ref{fig:XYpannel}, a 1D cross section of the density clearly shows the dip in the density at the center caused by the impurity.  We show curves for three different impurities, which we see minor differences in the density in the center with the strongest impurity creating the largest decrease in the density, as expected.  For the other impurities, both the depth and breadth of the density 'divot' is decreased.  Also, as is typical of condensates, far enough away from the impurity (further than the healing length) there is no distinction in the density profile between the different impurities, in this case at approximately 2 $\mu$m from the center.  In Figure \ref{fig:XYenergy}, we plot the self-energy (Eq. (\ref{selfE})) as a function of particle number for a few different impurity strengths.  Unsurprisingly, the self-energy is higher for larger interaction strengths.  With increasing particle number, it also increases and appears to scale linearly at large particle numbers.  We have added another curve corresponding to $^{159}$Tb, an as yet unexplored element with a strong magnetic dipole moment, in order to display another curve for strength dependence.  

Overall, the $xy$-confined system shows the expected isotropic response to the single impurity at the origin.  The impurity's effect is rather mild, as except for the inner $\sim$ 4\% of the trap, the density is indistinguishable from the case with no impurity.

\subsection{$xz$-plane}

\begin{figure}[!h]
\centering
\includegraphics[width=0.48\linewidth]{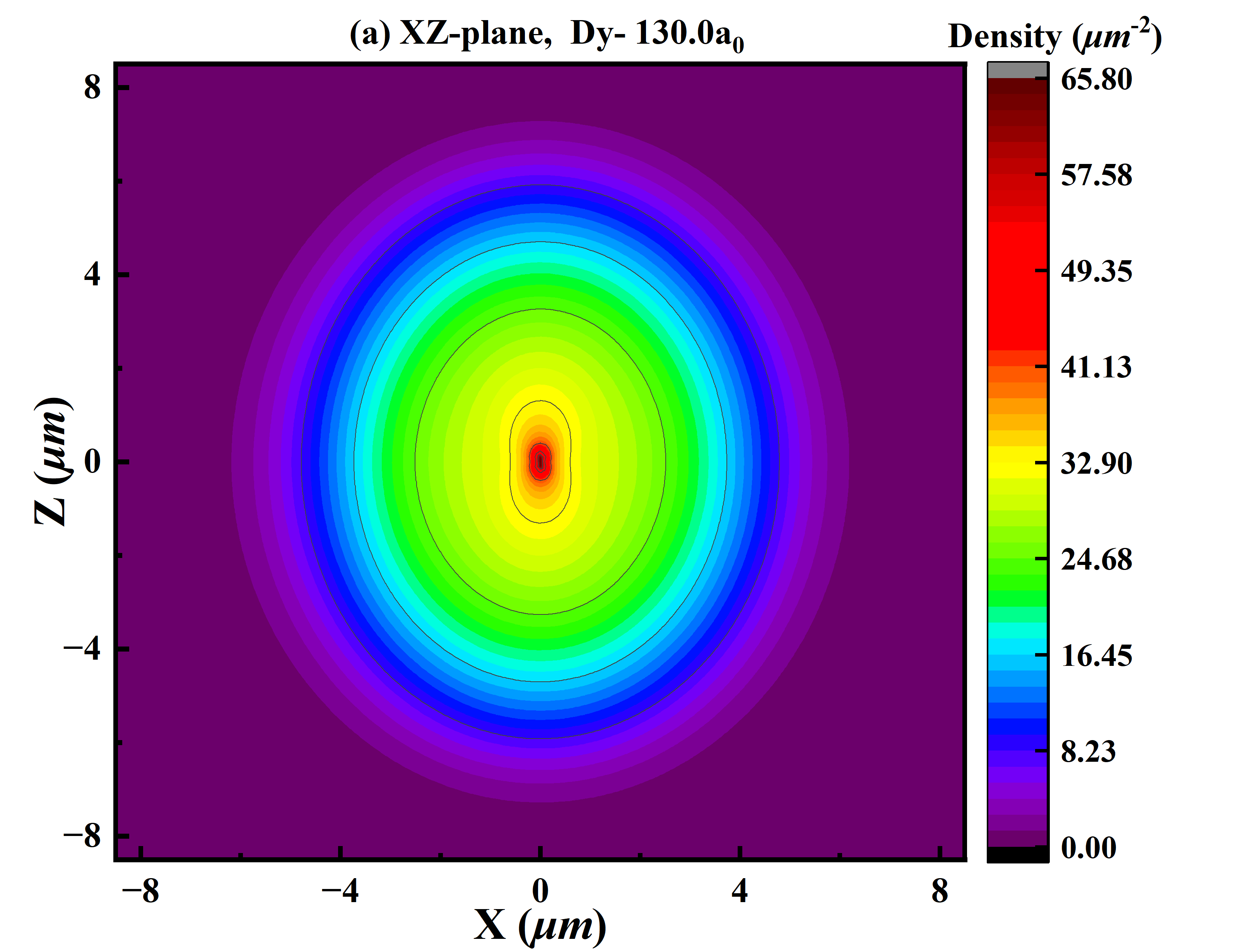}
\includegraphics[width=0.48\linewidth]{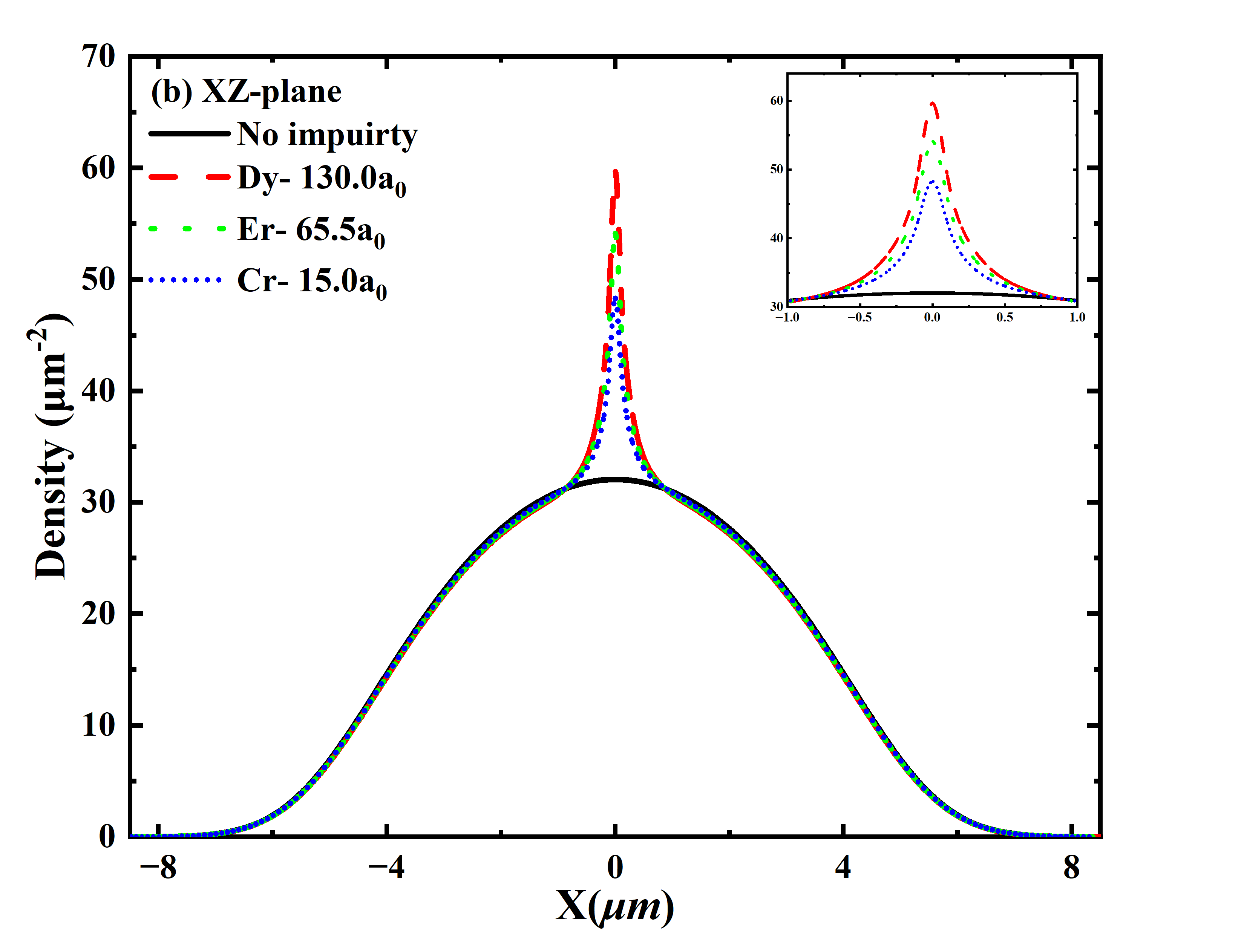}
\includegraphics[width=0.48\linewidth]{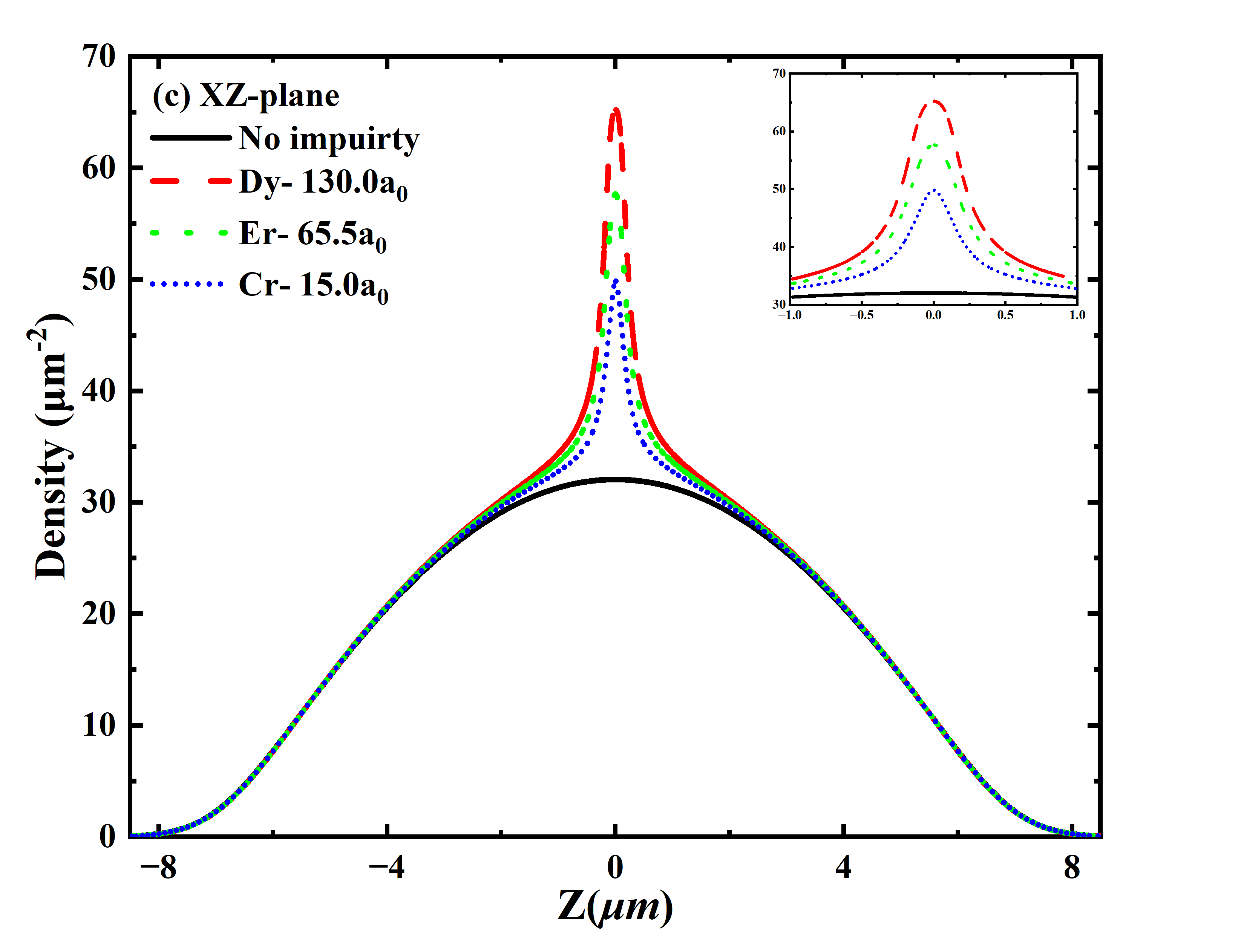}
\caption{\label{fig:XZpannel} 
Shown in Panel a) is a density contour plot for a dipolar gas consisting of 2000 Dy atoms confined in the $xz$-plane with dipole moments polarized parallel to the plane with an impurity Dy atom at the origin. Panels b) and c) show one dimensional density cross sections along the $x$- and $z$-axes, respectively, for three different impurities and the no impurity system.  Panels b) and c) also show insets showing a close up of the density profiles near to the origin.}
\end{figure}

If the particles are confined to a plane parallel to the dipolar axis, then $\theta_d$ in Eq. (\ref{dipole}) can take any value and we have an anisotropic system that, roughly speaking, should be attractive along the $z-$axis and repulsive along the $x-$axis.  The density contours of this system can be seen in panel a) of Figure \ref{fig:XZpannel}.  The shape is no longer circular as the contours are elliptical with the major axis along the $z-$axis.  The dipoles preferentially line up along the $z$-axis in order to keep their preferred head-to-tail configuration.  Since the trap is isotropic, this is the natural shape caused by the interactions within the gas.  This can also be seen in panels b) and c) of Figure \ref{fig:XZpannel} which show the 1D density cross sections along the $x$- and $z$-axes, respectively.  Looking at the spatial extent of these cross sections, one clearly sees that the $z$ density extends to larger values from the origin than along the $x$-axis. The overall extent is much smaller in the $xz$ case, which can be seen from the boundaries of the plots. The repulsive character of the $xy$ system forces the particles to spread out more and thus form a more dilute system overall, whereas the attraction present in the $xz$ and the large concentration in the center caused by the impurity means the total system's size is smaller in the $xz$ case.  The density of the $xz$ system does not drop to what was the peak density in the $xy$ system until about 4 $\mu$m from the origin.   

The effect of the impurity is to cause a spike in density at the very center, a spike which is wider in the $z$-direction than in $x$ (shown clearly in the insets of Figures \ref{fig:XZpannel}b and \ref{fig:XZpannel}c).  This is in contrast with the $xy$ confinement, where there was only repulsion and thus, a divot in the center.  The spike, however, is larger in magnitude to the no impurity density compared to the size of the divot, although in both cases they are around a factor of two relative to the pure system.  Still, far away from the impurity the density returns to the no impurity value. 

\begin{figure}[!h]
\centering
\includegraphics[width=0.48\linewidth]{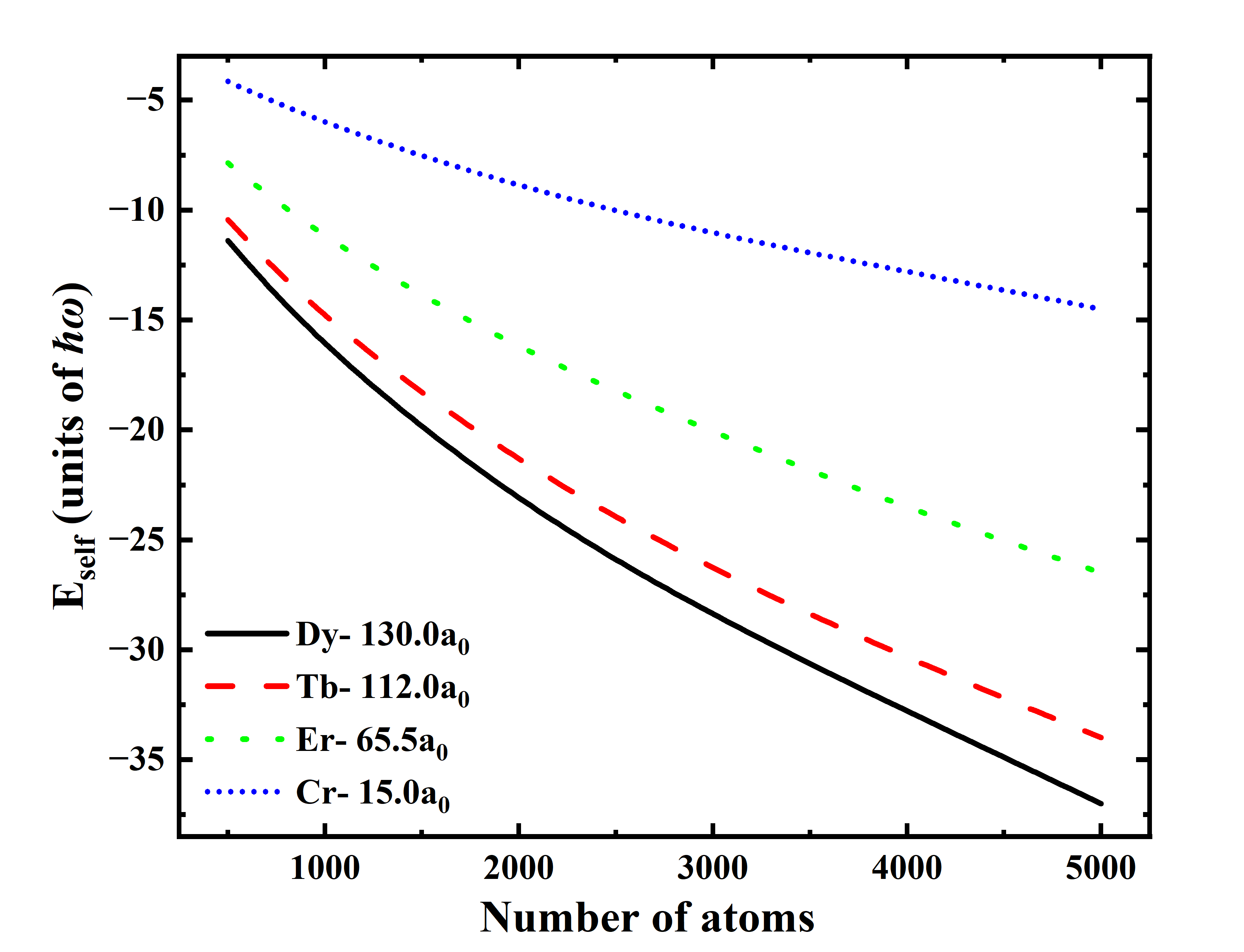}
\caption{\label{fig:XZenergy} Self-energy, Eq. (\ref{selfE}), of an impurity implanted into a dipolar gas confined to the $xz$-plane as a function of the number of particles in the dipolar gas for four different impurities.}
\end{figure}

Figure \ref{fig:XZenergy} shows the self-energy as a function of particle number.  Here, the self-energies are negative as the presence of the impurity adds attraction to the system and lowers the energy.  Its effect is greater with more particles present but overall behaves in a similar way to the $xy$-plane case, just attractive instead of repulsive.  However, it is not simply a reflection over the horizontal axis, as the self-energies for $xz$ are more negative than the $xy$ ones are positive.  As stated previously, the spike in density caused by the attractive impurity is larger than the density cavity caused by the repulsive impurity, so the impurtiy is affecting more of the bulk gas atoms and thus has a greater self-energy.  

\subsection{Anisotropic trap}

Experimenters have control over the geometry of their traps, and thus, the harmonic trap can be modified to make it anisotropic.  If the deformation is extreme, the system becomes quasi one-dimensional. Note we do not show results in the $xy$ confinement case, as the deformation did not change the self-energy.  The overall energy of the system did increase with increasing deformation, but these differences in particle distribution were far away from the center of the trap and therefore the self-energy remained constant.

For the $xz$ confinement, the trap potential is now 
\begin{equation}
    V_{trap}(x, z)=\frac{1}{2}m\left[\omega^2\left(b^2x^2 + a^2z^2\right)+\omega_{3D}^2y^2\right],
    \label{dftrap}
\end{equation}
where the parameters $a$ and $b$ are adjusted such that the area of the trap remains constant.  This means that $\sqrt{ab}=$ constant.  In order to quantify the deformation, we introduce the parameter $\varepsilon = a/b-b/a$.  For an isotropic case, $\varepsilon$ will be zero, then will be positive for a trap deformed along the $x$-axis and negative for a trap deformed along the $z$-axis.  We should also mention that the scenario of having an impurity pinned to the center of the trap by an external laser is most appropriate for this section, and would not make sense for a mobile impurity. 

\begin{figure}[!h]
\centering
\includegraphics[width=0.45\linewidth]{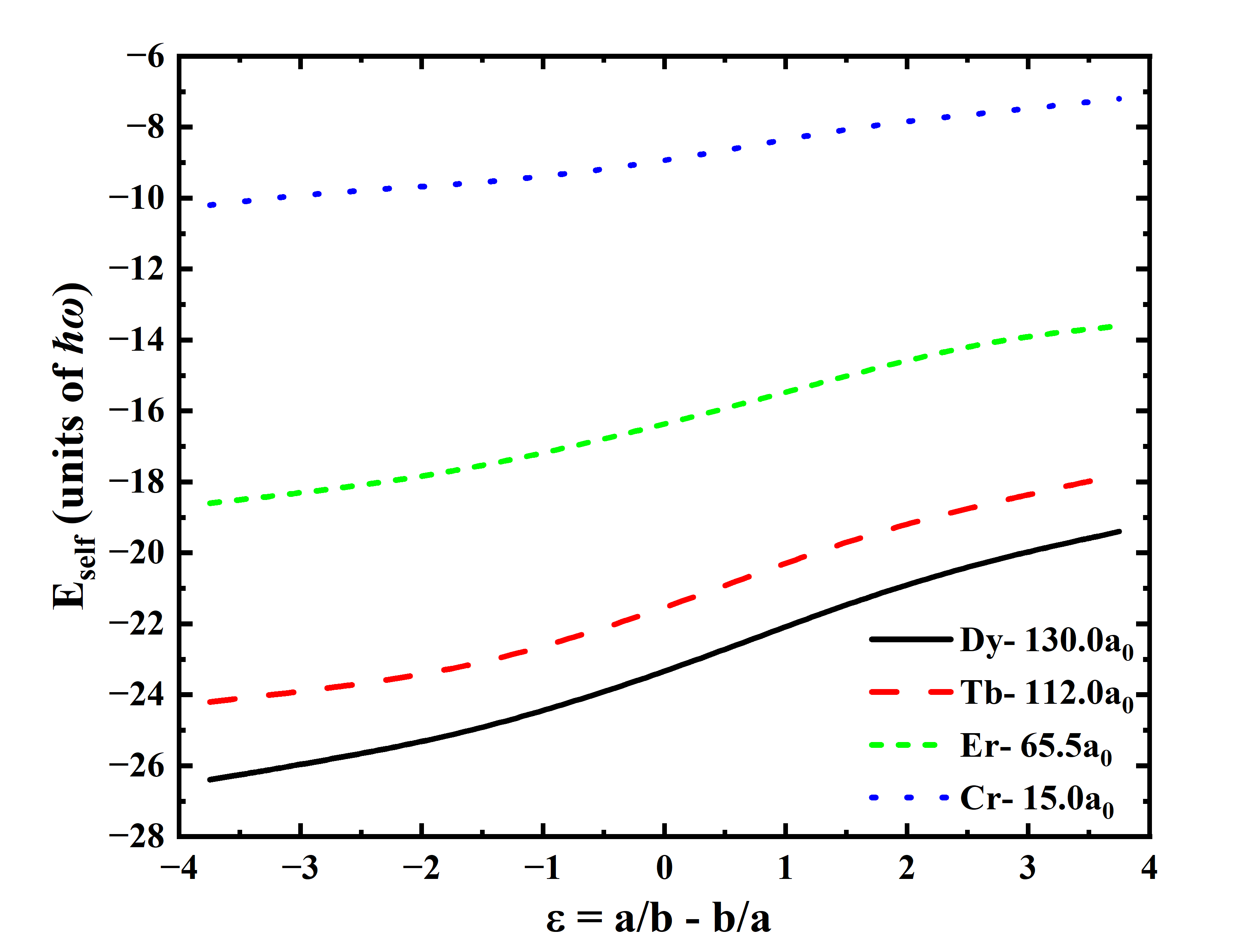}
\caption{\label{fig:XZ_deform} Self-energy plotted as a function of the deformation $\varepsilon$ for four different impurities.  These plots were obtained for a system of 2000 Dy atoms confined in the $xz$-plane with their dipole moments aligned along the $z$-axis.}
\end{figure}

\begin{figure}[!h]
\centering
\includegraphics[width=0.48\linewidth]{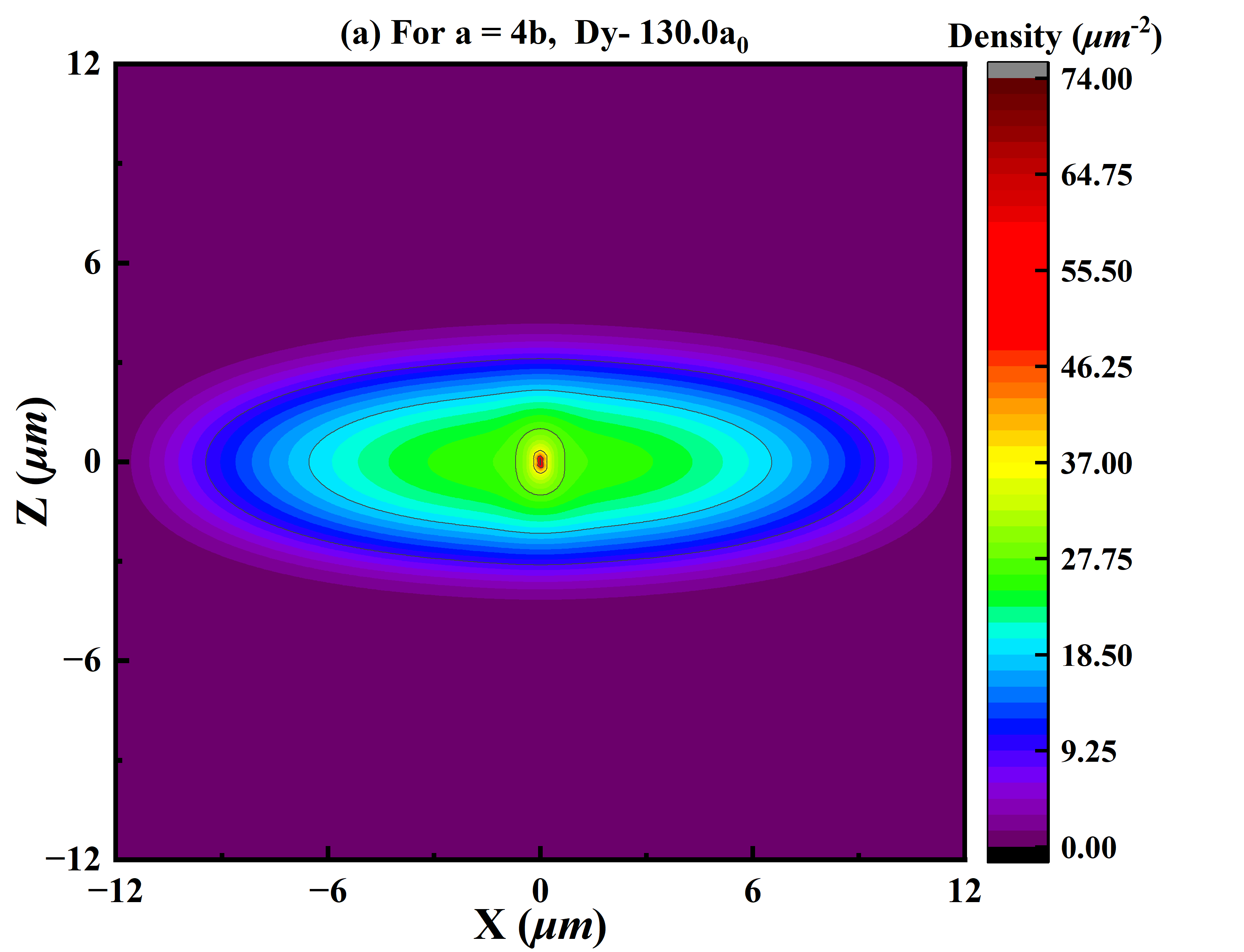}
\includegraphics[width=0.48\linewidth]{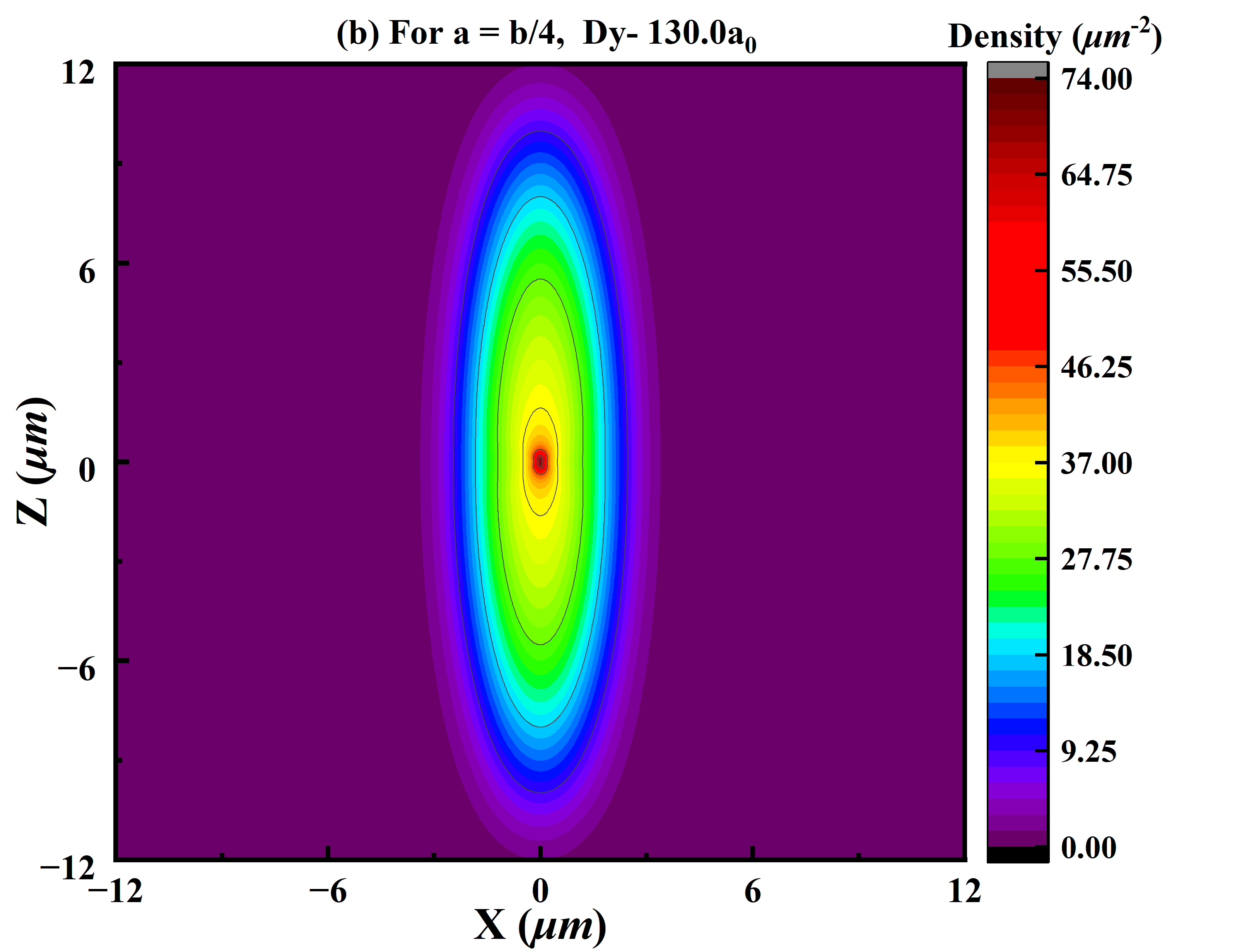}
\caption{\label{fig:XZ_defdens} Density contour plots for a system of 2000 Dy atoms confined in the $xz$-plane.  Here, the confining harmonic trap has been deformed as described in Eq. (\ref{dftrap}), and these results are for the most extreme deformations we considered which are $\varepsilon=\pm 3.75$.}
\end{figure}

The self-energy of the $xz$-system as a function of $\varepsilon$ is shown in Figure \ref{fig:XZ_deform}.  This quantity will show the interplay between the impurity and the trapping potential, while being insensitive to the interplay between the trap and the bulk medium. As before, the magnitude of the self-energies increase with the strength of the impurity dipole strength. With deformation, when the trap is deformed in the orthogonal direction to the external field, the self-energy increases, whereas when the trap deformation is parallel to the external field, it decreases. This makes sense as the perpendicular deformation forces more particles into the horizontal direction where dipoles will be in their repulsive orientation, including in the vicinity of the impurity and thus the difference in self-energies.  One can see this in the densities as well shown in Figure \ref{fig:XZ_defdens}.  In the upper panel, where $a=4b$ or $\varepsilon=3.75$, the contour lines closest to the impurities are elongated along the $z$-axis before changing to match the elongation of the trap further away.  The lines are not perfect elipses even further away from the origin.  They protrude slightly outward in the $z$-direction, reflecting the preferred alignment of the dipoles.  In the lower panel, the contours all match as the trap is elongated in the preferred direction of the gas.  Like in the upper panel, there is a slight deviation from a purely elliptical shape, where in this instance the contour lines slightly pinch inward along the $x$-axis.

\subsection{Time dynamics}

\begin{figure}[!h]
\centering
\includegraphics[width=0.48\linewidth]{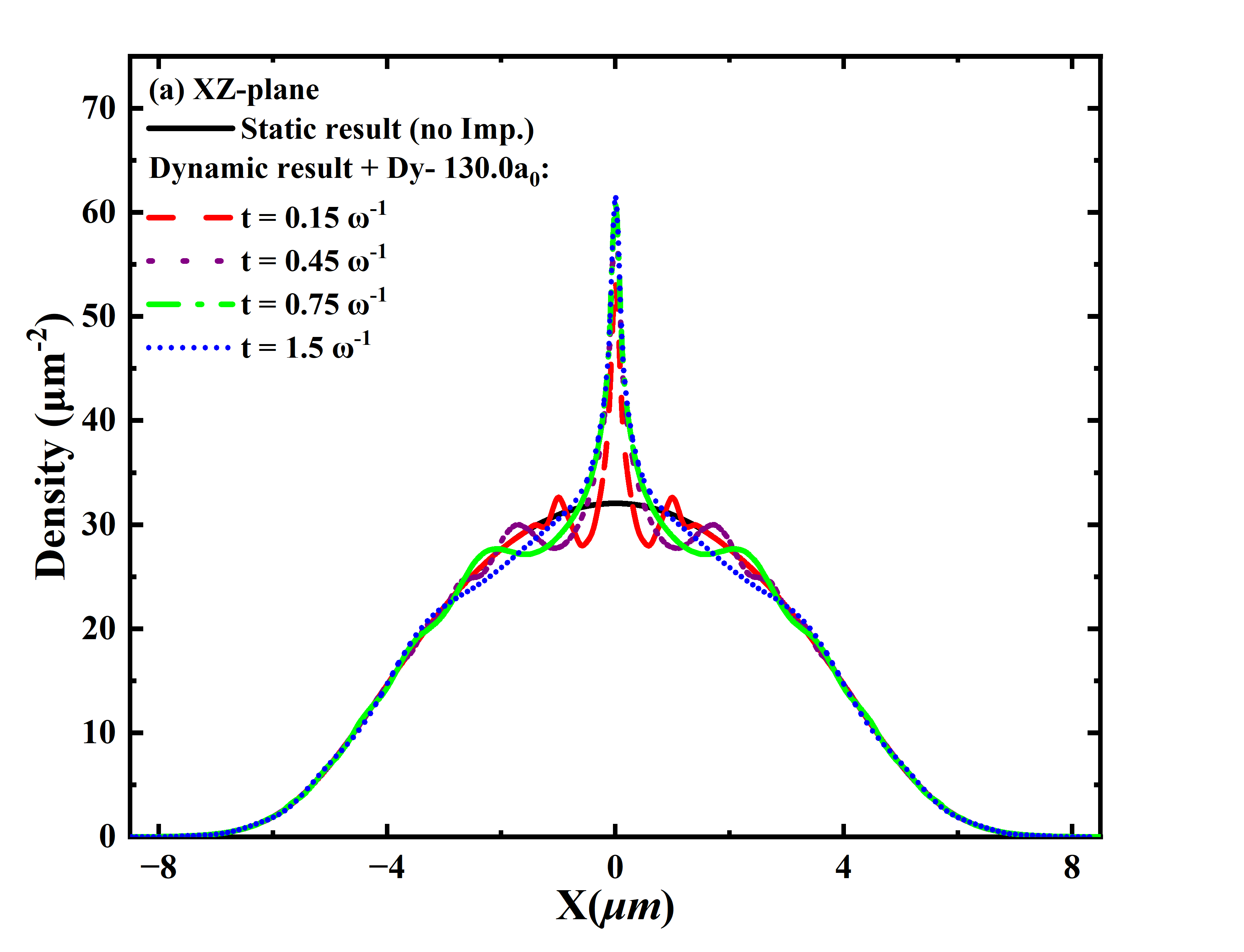}
\includegraphics[width=0.48\linewidth]{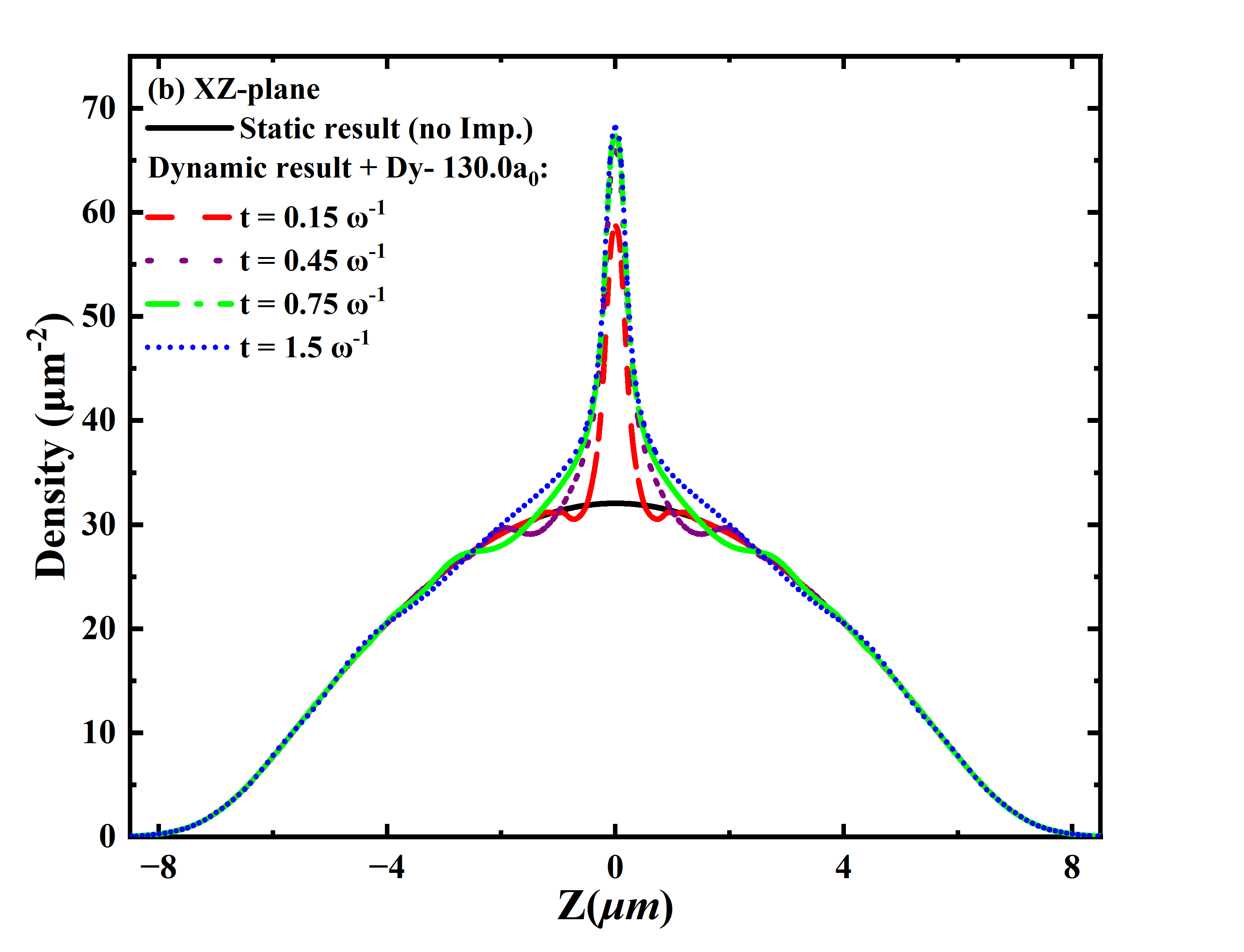}
\caption{  Density profiles along the $x$-axis (a) and the $z$-axis (b) for different amounts of time evolution.  At $t=0$, the dipolar BEC of 2000 Dy atoms confined in the $xz$ plane has a Dy impurity implanted at the origin.  The time unit, $\omega^{-1}$ corresponds to about 2.5 ms.  }\label{fig:XZ_time}
\end{figure}

\begin{figure}[!h]
\centering
\includegraphics[width=0.48\linewidth]{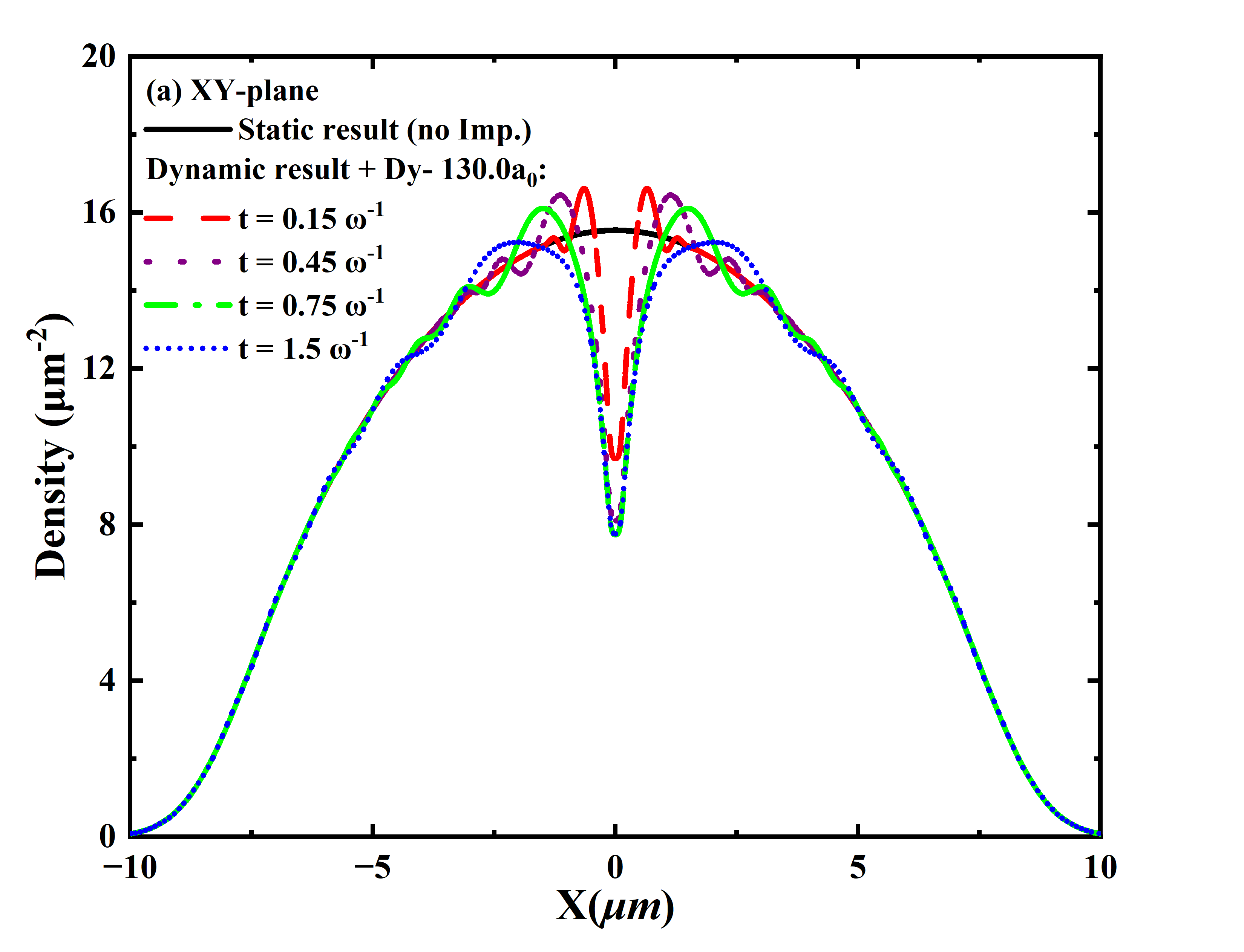}
\caption{  Density profile along the $x$-axis for different amounts of time evolution.  For the same gas as in Figure \ref{fig:XZ_time}, only this gas is confined to the $xy$ plane.  Since the system is isotropic, the profile along the $y$-axis is identical. }\label{fig:XY_time}
\end{figure}

In order to understand how a dipolar gas reacts to the introduction of an impurity, we employ the real-time propagation of the split-step Crank-Nicolson method in our numerical calculations.  In these calculations, the dipolar gas without an impurity is our initial state, which we then propagate in real-time with the impurity potential.  The density results of these calculations are seen in Figure \ref{fig:XZ_time} for the $xz$-plane and Figure \ref{fig:XY_time} for the $xy$-plane.  Since the $xy$-plane results are isotropic, we only show results along the $x$-axis.  In the case of the $xz$ results, one can see at small times the density increasing in the center from the introduction of the impurity, but also causing ripples in the density that extend further out to distances beyond what was seen in the static results.  As the time increases, the central spike increases in height and the ripples dampen out as the profile approaches the static result.  Broadly, the behavior is the same along both the $x$ and $z$ axes, however they are not identical.  Due to the larger extent in the $z$ direction, the ripples appear smaller in amplitude, and, as in the static case, the central maximum is wider.  

In the case of the $xy$-plane system (Figure \ref{fig:XY_time}), the depth and width of the central density minimum increases with time.  The ripples increase in amplitude towards the center as they decrease over time.  As in the $xz$ case, the density ripples extend far outside the region where the static density profile differed from the no impurity profile (about $\pm$2 $\mu$m).

\section{Conclusion}

In this paper we have shown density and self-energy results for a dipolar impurity immersed in a 2D dipolar BEC for two different geometries.  We have found that the impurity distorts the gas in its immediate environment while mostly leaving the gas far away unchanged.  Our dynamic results show, however, that when an impurity is first introduced it can cause density fluctuations at a much greater distance.  We have primarily used impurity strengths that are experimentally relevant and hope that our results provide some impetus for experimental exploration of these systems.  

Some related systems that could be studied further include adding an additional impurity or impurities.  Not only would they add further distortion to the BEC, but the impurities themselves would interact via the exchange of elementary excitations of the medium\cite{paredes}.  One could also imagine tilting the confinement to orientations between the $xy$- and $xz$-plane extremes.  One might expect interesting behavior at the so-called 'magic angle'$=\cos^{-1}(1/\sqrt{3})\approx54.74^\circ$ where the dipolar interaction vanishes in one direction, but the interaction in the perpendicular direction would remain dipolar.

\section{Acknowledgements}
The authors acknowledge that this material is based upon work supported by the National Science Foundation/EPSCoR RII Track-1: Emergent Quantum Materials and Technologies (EQUATE), Award OIA-2044049.

\newpage

\bibliographystyle{apsrev4-1}
\bibliography{ref}

\end{document}